# Ultrafast photoinduced enhancement of nonlinear optical response in 15-atom gold clusters on indium tin oxide conducting film

Sunil Kumar,[1,§] E. S. Shibu,[2] T. Pradeep,[2] and A. K. Sood[1,*]

[1]*Department of Physics, Indian Institute of Science, Bangalore-560012, India*
[2]*Department of Chemistry, Indian Institute of Technology Madras, Chennai 600036, India*
*Electronic address:* [§]sunilvdabral@gmail.com; *asood@physics.iisc.ernet.in

**Abstract:** We show that the third order optical nonlinearity of 15-atom gold clusters is significantly enhanced when in contact with indium tin oxide (ITO) conducting film. Open and close aperture z-scan experiments together with non-degenerate pump-probe differential transmission experiments were done using 80 fs laser pulses centered at 395 nm and 790 nm on gold clusters encased inside cyclodextrin cavities. We show that two photon absorption coefficient is enhanced by an order of magnitude as compared to that when the clusters are on pristine glass plate. The enhancement for the nonlinear optical refraction coefficient is ~3 times. The photo-induced excited state absorption using pump-probe experiments at pump wavelength of 395 nm and probe at 790 nm also show an enhancement by an order of magnitude. These results attributed to the excited state energy transfer in the coupled gold cluster-ITO system are different from the enhancement seen so far in charge donor-acceptor complexes and nanoparticle-conjugate polymer composites.

## 1. Introduction

Metal nanoparticles have attracted great deal of interest due to their fascinating optical and optoelectronic properties [1, 2]. The surface plasmon resonance in metal nanoparticles is not only responsible for their linear optical properties but also governs the nonlinear optical effects. Near the surface plasmon resonance, metal nanoparticles exhibit immensely enhanced optical nonlinearities as compared to their bulk counterparts [3-6]. Smaller metal nanoparticles, namely, the nanoclusters containing a few atoms are a new class of materials with characteristics in between those of atoms and nanoparticles [7-9]. These are too small to be described as nanoparticles and their electronic structure is characterized by well defined singlet and triplet molecular states rather than a continuous density of states. The optical absorption spectra of such clusters lack the surface plasmon resonance (usually seen for particles bigger than 2 nm) and shows a distinct absorption onset at the electronic gap between the highest occupied molecular orbital (HOMO) and the lowest unoccupied molecular orbital (LUMO). These have been shown to exhibit several novel properties [10-12]. A few ultrafast studies on gold clusters have been reported in the literature [13-15]. On bigger gold clusters containing 140 atoms, Thomas *et al.,* [13] observed optical limiting using nanosecond laser pulses at 532 nm, suggesting its origin to fifth order nonlinearity, $\chi^{(5)}$ with a three photon absorption coefficient of ~2.9x10$^{-15}$ cm$^3$/W$^2$. Such a conclusion was drawn from fitting the z-scan data with equations incorporating two- and three-photon absorption coefficients. Photoexcited electron relaxation dynamics in smaller gold nanoclusters containing 28 atoms surrounded by glutathione molecules was reported [14] to show bi-exponential decay with a fast sub-picosecond and another slow (ns) time-constant, independent of both the laser photon energy and the pump-fluence.

In this paper, we have investigated nonlinear optical response of 15-atom gold clusters ($Au_{15}$) using femtosecond nonlinear transmission z-scan and transient differential transmission pump-probe spectroscopy. It is seen that the nonlinear response is immensely enhanced for the clusters deposited on an indium-tin-oxide (ITO) metal film as compared with that on $SiO_2$ glass plate. We have systematically carried out femtosecond z-scan experiments at 395 nm and time-resolved non-degenerate pump-probe experiments (395 nm pump and 790 nm probe) on the gold-clusters deposited on glass plate ($Au_{15}$-$SiO_2$) and on ITO ($Au_{15}$-ITO) as well as separately on bare ITO and glass plates. The optical limiting performance of the $Au_{15}$-ITO system is higher by almost an order than that of the $Au_{15}$-$SiO_2$ system. Concurrently, excited state absorption from transient transmission measurements is also enhanced by the same order. From our experimental results we have estimated the two photon absorption coefficient β of ~0.3 (3.0) cm/GW and the nonlinear refraction coefficient γ of ~2x10$^{-5}$ (6x10$^{-5}$) cm$^2$/GW for the $Au_{15}$-$SiO_2$ ($Au_{15}$-ITO) system. We find that the excited state relaxation dynamics involves two-step decay with fast time-constant of ~700 fs and the other much slower with time constant ~1 ns similar to that reported for 28 atom gold clusters [14]. In charge donor-acceptor complexes, photoinduced electron or energy transfer has been observed to significantly improve the optical limiting [16-19]. We suggest a similar mechanism in our case involving ultrafast intersystem excitation transfer from the photoexcited $Au_{15}$ to the ITO. Since the ground state single photon absorption at 790 nm and 395 nm are found to be similar for both the $Au_{15}$-ITO and $Au_{15}$-$SiO_2$ samples, the electronic coupling via intersystem crossing occurs in the excited state of the coupled system. Enhanced excited state absorption in the $Au_{15}$-ITO sample as compared to $Au_{15}$-$SiO_2$ sample from our transient differential transmission measurements justifies the intersystem crossing in the excited state of the cluster-ITO coupled system.

## 2. Experimental details

Stable $Au_{15}$ nanoclusters were synthesized involving processes like core-etching of larger clusters and simultaneous trapping of the clusters inside cyclodextrin (CD) cavities [10] shown schematically in inset of Fig. 1. The decorating glutathione (GSH) and CD molecules are optically transparent in the 200-1000 nm window. The absorption spectrum of the gold clusters is shown in Fig. 1 where the experimentally obtained wavelength (W) dependent intensity I(W) has been converted into energy (E) dependent intensity by dividing by the popular Jacobian factor ∂E/∂W. We note that the absorption profile of the clusters does not show surface plasmon resonance (SPR). Instead, molecule-like characteristic features at about 318 nm, 458 nm and 580 nm are clearly visible. The distinct absorption onset occurs near 710 nm which indicates an energy gap of 1.75 eV between the highest occupied molecular orbital (HOMO) and the lowest unoccupied molecular orbital (LUMO) gap. We may note that this is much larger than the theoretically predicted HOMO-LUMO gap (~0.2 eV) in neutral $Au_{15}$ clusters [20].

For our present study, 80 mg of the nanoclusters taken in a powder form with 160 µL water (50% by weight) forms a dense hydrogel matrix. The hydrogel of thickness ~100 micron was sandwiched between a glass plate (~500 micron thick) and a coverslip (~100 micron thick) making the $Au_{15}$-$SiO_2$ sample and between an ITO thin film-coated glass plate and a cover slip to make the $Au_{15}$-ITO sample. The commercial ITO plates having film thickness of ~100 nm and a sheet resistance of ~10 ohm per square area were used. When excited with 395 nm laser pulses, a yellowish glow is seen at the back of the samples. We used a color glass wavelength filter to avoid fluorescence from the nanoclusters from reaching the photo-detector in all our experiments described below.

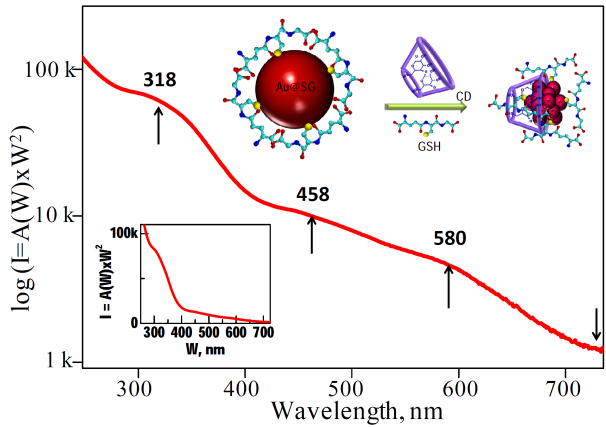

Fig. 1. UV-Vis absorption spectrum of $Au_{15}$ clusters inside cyclodextrin (CD) cavities plotted as the natural logarithm of the Jacobian factor. Well defined absorption features related to molecular character of the clusters are marked with upward arrows. Lower inset shows a plot of the absorption spectrum on a linear-linear scale. Upper inset shows the schematic illustration of CD-assisted one-pot synthesis of the $Au_{15}$ clusters via the core etching reaction [10]. GSH indicates glutathione used as the core-etching agent and as the ligand to protect the $Au_{15}$ core.

Nonlinear transmission experiments were carried out at 395 nm (3.15 eV) using z-scan method in both open aperture (OA) and closed aperture (CA) configurations. The fundamental laser beam at 790 nm was taken from a 1-kHz repetition rate amplifier system (Spitfire, Spectra Physics) while the second harmonic at 395 nm was generated using a 500 micron thick beta barium borate crystal. A 5 mm circular diameter beam was selected using an iris placed before the z-scan setup. The incident power on the sample was varied from ~2 MW/cm$^2$ (far from focal point) to ~70 GW/cm$^2$ (at the focal point, z = 0) using a focusing lens. In the OA case, all the transmitted intensity was collected using a collection lens, whereas in the CA case an aperture with circular diameter of 2.5 mm was placed on the photodiode. We also carried out transient transmission measurements using pump pulses centered at 395 nm and probe pulses centered at 790 nm in usual non-collinear pump-probe geometry. At the sample point, the laser pulse-duration was measured to be ~80 fs. The pump-fluence was ~16 µJ/cm$^2$, whereas that of the probe was ~1 µJ/cm$^2$. We must mention that z-scan and degenerate pump-probe experiments at 790 nm did not give measurable results from the gold cluster samples even upto higher fluences (~1 mJ/cm$^2$) clearly indicating that the nonlinear response is negligibly small at 790 nm. All our experimental results reported here were repeatable on the same sample suggesting that the nanoclusters are very stable even after laser photoexcitation. This is possibly due to the CD encapsulation which is known to enhance chemical stability [10].

## 3. Results

In Figs. 2(a) and 2(b) we have presented the normalized transmittance for the $Au_{15}$-ITO and $Au_{15}$-SiO$_2$ samples as obtained from z-scan in both the OA and CA configurations. The signal from the OA z-scan shows optical limiting, i.e., reduction in transmission as the input beam intensity is increased, whereas the signal from CA z-scan shows a positive refractive nonlinearity (self-focusing effects), i.e., decrease in the transmitted intensity due to refraction as the sample approaches the focal point (z = 0) followed by an increase in intensity as it moves away from the focal point and towards the detector. We note from these results that the magnitude of change in transmission near the focal point is about 10 times larger for the $Au_{15}$-ITO sample as compared to the $Au_{15}$-SiO$_2$ sample. We may point out that the optical limiting behavior seen in femtosecond experiments does not arise from thermal effects-induced transient refractive index changes. In general, scattering losses are dominant at the nanosecond or higher time-scales. However, in the femtosecond regime, fast electronic effects are the principal contributors to nonlinear optical susceptibility such as $\chi^{(3)}$. Therefore, the observed nonlinear responses of our samples in Fig. 2 are purely electronic in nature.

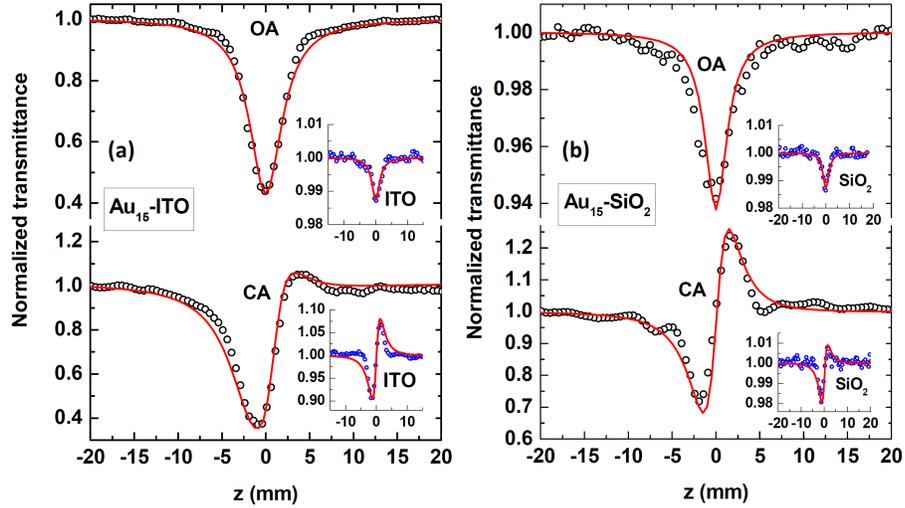

Fig. 2. Nonlinear transmission results measured using z-scan method in both OA and CA configurations taken at an excitation wavelength of 395 nm (3.15 eV). Normalized transmittance of $Au_{15}$ clusters in contact with (a) ITO-coated glass plate, and (b) bare glass plate. The insets in (a) and (b) show results for bare ITO and glass plates alone. The continuous curves are numerical fits to retrieve the two-photon absorption and nonlinear refraction coefficients as given in Table 1.

Our first approach here is to estimate the nonlinear absorption and refraction coefficients of our samples by fitting the experimental data in Fig. 2. Limiting ourselves up to third order nonlinear effects, the effective absorption and refraction coefficients can be written as $\alpha = \alpha_0 + \beta I$ and $n = n_0 + \gamma I$, where $\alpha_0$ and $n_0$ are the linear absorption coefficient and refractive index, $\beta$ is the two photon absorption coefficient, $\gamma$ is the nonlinear refractive index and $I$ is the input intensity. Applying the z-scan theory [21, 22] for the open aperture and close aperture configurations, we have fitted the experimental data in Fig. 2 (fits are shown by continuous lines). For the fitting we take $n_0 \sim 1.5$ and $\alpha_0 \sim 1.5$ cm$^{-1}$ as measured separately from linear transmission of both the Au$_{15}$-ITO and Au$_{15}$-SiO$_2$ samples. We can clearly see that only third order nonlinear terms are sufficient to obtain good fits to our nonlinear transmission data. Thus obtained nonlinear parameters have been given in Table 1. We note from Table 1 that the value of $\beta$ ($\gamma$) for the Au$_{15}$-ITO sample is about 10 (3) times higher than that for the Au$_{15}$-SiO$_2$ sample.

For a reference, we performed the OA and CA z-scan experiments on bare ITO and glass plates as well for which the data are shown in the insets of Figs. 2a and 2b, and the corresponding nonlinear coefficients given in Table 1. The values of $\beta$ and $\gamma$ for the bare ITO and glass plates are negligibly small (Table 1) as compared to those for the Au$_{15}$-ITO or Au$_{15}$-SiO$_2$ samples. Previously, using 200 fs laser pulses centered at 720 nm, nonlinear optical coefficients of ITO films of a typical thickness ~120 nm and carrier concentration of ~5 x 10$^{20}$ cm$^{-3}$ have been reported [23] to be $\beta \sim 0.1$ cm/GW and $\gamma \sim 4 \times 10^{-5}$ cm$^2$/GW whereas these parameters for SiO$_2$ are atleast two orders smaller. However, our present z-scan experiments using 80 fs laser pulses centered at 395 nm show that for both ITO and SiO$_2$, $\beta$ and $\gamma$ are roughly of the same order (Table 1) as well as smaller than those reported at 720 nm [23].

The results for the transient differential transmission ($\Delta T/T$) from pump-probe spectroscopy using 395 nm pump and 790 nm probe are presented in Fig. 3. The negative sign of the transient differential transmission signal is in contrast to the positive signal (photo-bleaching) usually seen for larger colloidal gold nanoparticles at the surface plasmon resonance. Here, the signal corresponds to excited-state absorption and not the ground state bleaching [14]. The magnitude of the signal at time delay, $t = 0$ is a direct consequence of the magnitude of the excited state absorption in the gold clusters at the probe wavelength of 790 nm. We can clearly see from Fig. 3a that the excited state absorption in the gold clusters is enhanced by a factor of ~8 for the clusters deposited on ITO as compared to those on the glass plate. This enhancement is similar to that observed for the optical limiting performance of the Au$_{15}$-ITO sample. In addition, ITO alone does not show any pump-probe signal, although at the highest experimental pump-fluence of ~1 mJ/cm$^2$ we see a small and symmetric dip in the differential transmission around the zero delay (inset of Fig. 3a), which is related to coherent artifact.

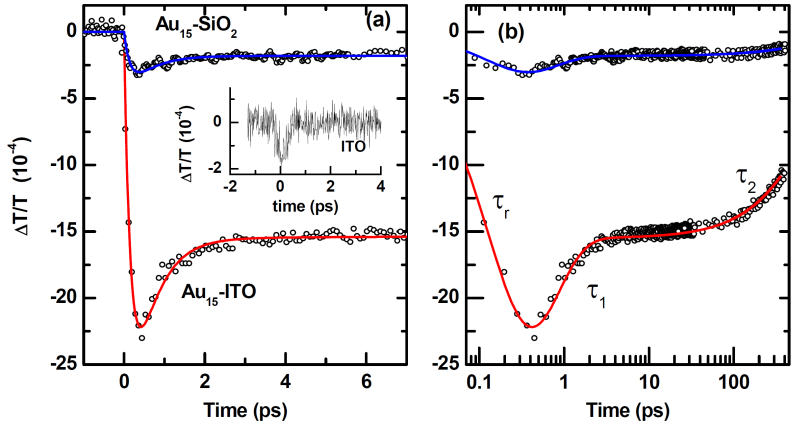

Fig. 3. Transient differential transmission data from non-degenerate pump-probe spectroscopy of Au$_{15}$-ITO and Au$_{15}$-SiO$_2$ samples using 395 nm pump and 790 nm probe, shown in (a) linear-linear plot, and (b) log-linear plot. Continuous red and blue curves are fits using bi-exponentially decaying function with fast time-constant $\tau_1$ and slow time-constant, $\tau_2$ along with a rising contribution with rise time $\tau_r$. The inset in (a) is the result on an ITO plate using the highest experimental pump-fluence of ~1 mJ/cm$^2$.

The life-time of carriers in the excited state has been determined by fitting a bi-exponentially decaying function to the photo-induced absorption in Fig. 3 where the data in Fig. 3b has been plotted on a linear-log plot to clearly show the initial buildup time $\tau_r$, fast time constant $\tau_1$ and the slow time constant $\tau_2$. For both the Au$_{15}$-ITO and Au$_{15}$-SiO$_2$ samples we have found $\tau_r \sim 200$ fs, $\tau_1 \sim 700$ fs and $\tau_2 \sim 1$ ns. The fast decay is attributed to the initial carrier relaxation from optically coupled states to the LUMO level followed by the slow decay via radiative and/or nonradiative processes from the LUMO level to the ground state. We find (data not shown) that the magnitude of the pump-induced absorption linearly increases with the excitation power (pump-fluence) indicating one-photon absorption processes at 3.15 eV. The carrier life-times remain pump-fluence independent. The photoexcited electron relaxation dynamics in our $Au_{15}$ clusters is similar to that reported previously for 28-atom gold clusters surrounded by glutathione molecules [14].

The magnitude of the photoinduced excited state absorption at time delay t = 0 (Fig. 3) can be used to quantify the cascaded two-photon absorption process (one photon of 395 nm and the other of 790 nm) by estimating the corresponding nonlinear absorption coefficient $\beta_{(2)}$ via the relation $\left|\Delta T/T\right|_{t=0} = \beta_{(2)}\sqrt{I_{pump} \times I_{probe}}$. The estimated value is $\beta_{(2)} \sim 0.05$ cm/GW for $Au_{15}$-ITO and 8 times smaller value for $Au_{15}$-$SiO_2$.

Table 1. Nonlinear optical coefficients obtained from z-scan measurements at 395 nm (3.15 eV). Here β is two-photon absorption coefficient, γ is nonlinear refraction coefficient and $Re\chi^{(3)}$ and $Im\chi^{(3)}$ represent the real and imaginary parts of the third-order optical susceptibility, $\chi^{(3)}$.

| Sample | β (cm/GW) | γ (cm²/GW) | $Re\chi^{(3)}$ (esu) | $Im\chi^{(3)}$ (esu) |
|---|---|---|---|---|
| $SiO_2$ | $5 \times 10^{-3}$ | $6 \times 10^{-7}$ | $8.5 \times 10^{-15}$ | $1.4 \times 10^{-15}$ |
| ITO | $8 \times 10^{-3}$ | $6 \times 10^{-7}$ | $8.5 \times 10^{-15}$ | $2.2 \times 10^{-15}$ |
| $Au_{15}$-ITO | 3.0 | $6 \times 10^{-5}$ | $8.5 \times 10^{-13}$ | $8.4 \times 10^{-13}$ |
| $Au_{15}$-$SiO_2$ | 0.3 | $2 \times 10^{-5}$ | $2.8 \times 10^{-13}$ | $8.4 \times 10^{-14}$ |

**4. Discussion**

Due to very small optical nonlinearity shown by the $SiO_2$ glass plate alone, the $Au_{15}$-$SiO_2$ sample can be regarded as just $Au_{15}$. Enhancement in the excited-state absorption observed from our pump-probe measurements and third-order optical susceptibility using z-scan in the $Au_{15}$-ITO sample is due to intersystem excitation transfer. Two mechanisms, the energy transfer and the electron transfer can lead to intersystem excitation transfer [18]. Resonance energy transfer between charged species in the donor-acceptor pairs is well known to significantly enhance the linear and nonlinear optical performance of the coupled pair systems [18, 19]. For resonant energy transfer, good overlap between the absorption spectrum of one and the emission spectrum of the other charged constituent of the binary system is required. The internal electron transfer mechanism requires sub-ps carrier diffusion times [18]. The gold nanoparticles, usually negatively charged [24] act as good electron or energy acceptor materials [18, 19]. Optical limiting in the gold nanoparticles is significantly enhanced when they are embedded in charged polymers [18, 19, 25]. Out of the two mechanisms, resonance energy transfer is believed to play a major role for the bigger gold nanoparticles due to sufficient overlap between the polymer emission and the nanoparticles absorption spectra. On the other hand, for smaller nanoparticles (size < 2 nm), the spectral overlap is insufficient for energy transfer mechanism to contribute. Rather, electron transfer contributes to the enhanced optical performance of the composite system [18]. The above description is for the ground state intersystem crossing. In comparison, photoinduced intersystem crossing via excited state energy or electron transfer is yet to be investigated.

Conducting metal oxide, ITO is a heavily doped n-type degenerate semiconductor where $Sn^{4+}$ at $In^{3+}$ sites act as donors when the oxygen stoichiometry is properly controlled and behaves as a metal [26]. Thin films of ITO are transparent in near-infrared and visible regions as a consequence of the lack of interband transitions. The plasma frequency in thin films of ITO is of ~1 eV [26, 27]. Moreover, the absorption spectra of ITO films suggest that the indirect and direct band gap energies lie around 2.4 eV (515 nm) and 3.6 eV (345 nm), respectively [28]. The $Au_{15}$ clusters are anionic in character [29]. Therefore, the gold clusters studied by us can be electron acceptors. At first, it appears that in our $Au_{15}$-ITO coupled systems the gold clusters act as acceptor species and the ITO act as the donor counterpart where excited state electron or energy transfer via inter-system coupling (ISC) plays a role in the observed enhanced nonlinear response. However, we must note that since the clusters are protected with glutathione encased in CD cavities [10], the direct interaction between clusters and electron-donor ITO via electron transfer is less likely. Moreover, we find that the single photon absorption at 790 nm or 395 nm is the same for both the $Au_{15}$-$SiO_2$ and $Au_{15}$-ITO systems. We further note that the distance over which resonance energy transfer can happen between localized center and delocalized electronic states on a substrate can be larger as compared to two localized centers [30]. Therefore, we suggest that in our $Au_{15}$-ITO coupled system where excited states of the $Au_{15}$ clusters and the ITO show an overlap, photoinduced energy transfer is responsible for the observed nonlinear optical results.

We construct a physical picture as depicted in Fig. 4. Here, on the left side in the figure, we have schematically drawn the molecular energy level diagram of the gold clusters [31] considering the absorption spectrum given in Fig. 1. The HOMO-LUMO gap of ~1.75 eV is very close to our probe photon energy of 1.55 eV. Three molecular like levels have been drawn at energies (wavelengths) of 2.1 eV (580 nm), 2.7 eV (458 nm) and 3.9 eV (318 nm). Similarly, on the right side in Fig. 4, we have drawn schematically, the energy bands in ITO. In the $Au_{15}$-ITO coupled system, due to close proximity of the energy levels of the molecular $Au_{15}$ clusters and energy bands of ITO, excited state electronic-coupling can occur. In that case the electrons created in the excited state of $Au_{15}$ (marked by A) can cross-over into the ITO side which further absorb the 395 nm or 790 nm photons leading to enhanced combined two photon absorption in the coupled system. The electronic transitions following absorption of a photon of wavelength 395 nm or 790 nm are shown by vertical blue and red arrows in Fig. 4.

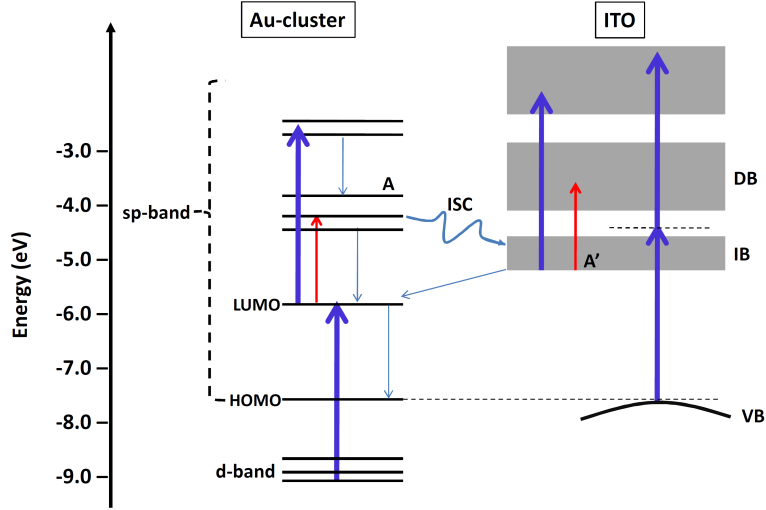

Fig. 4. Schematic of the electronic energy-diagram of $Au_{15}$ cluster coupled with the electronic energy-band diagram of ITO. The blue and red arrows represent the pump and probe photon energies. Possible excited-state intersystem coupling between $Au_{15}$ clusters and ITO is represented by the curved arrow. Some of the possible interstate relaxation processes are marked by thin downward arrows.

In a particular case, the optical limiting behavior of a composite system gets enhanced by coupling efficient two-photon absorption in one member with reverse saturable absorption or optical limiting in the other member [17], provided the emission from the first member following its two-photon absorption closely matches with the excited state absorption in the second member of the pair. We have a similar situation where the $Au_{15}$ and ITO individually show optical limiting at 395 nm while excited state absorption at 790 nm or 395 nm occurs in $Au_{15}$. The optical absorption spectra [10] as well as our pump-probe results on $Au_{15}$ clusters show that there is enough single photon absorption at 395 nm. The $Au_{15}$ clusters have a strong broad band emission centered at around 680 nm [10] but this radiation or the 395 nm can-not excite the ITO in its ground state. Moreover, in ITO alone the direct transition probability from its ground state (top of the valence band VB) to the indirect band (IB) at 2.4 eV is negligibly small as inferred from our experiments and the transition to the direct band at 3.6 eV does not occur at our photon energy of 3.15 eV. Since the enhancement in the excited state absorption (Fig. 3) is similar to that in the optical limiting (Fig. 2), it is natural to assume that the optical limiting in $Au_{15}$ at 395 nm is also a two-step cascaded single photon absorption process. Hence, we consider a two-step cascaded single photon absorption processes in the $Au_{15}$ clusters and direct two photon absorption via an intermediate virtual state (much smaller absorption cross-section) taking place in ITO at 395 nm (shown by two thick blue arrows in Fig. 4). After excited state electron transfer from $Au_{15}$ via intersystem electronic coupling, the electrons cross-over to the IB state of ITO which can readily absorb single photon of either 395 nm or 790 nm leading to the observed enhanced two photon absorption in z-scan or excited state absorption in pump-probe experiments on the $Au_{15}$-ITO system.

In the coupled $Au_{15}$-ITO system, the excited state absorption of a 395 nm or 790 nm photon takes place from the states marked as LUMO (or A' following the intersystem crossing) and the probe monitors the carrier life time in these states as a function of pump and probe delay time. The fast component with time constant ~700 fs can be related to the interstate relaxation of the carriers (marked by downward arrows in Fig. 4) until they reach the lowest energy excited state from which they ultimately relax to the ground state *via* radiative and/or nonradiative emission within the longer time of $\tau_2$ ~1 ns as observed in the experiments.

From our z-scan results in Fig. 2 we note that the nonlinear threshold (incident intensity where the transmission begins to decrease) is much smaller at ~10 GW/cm$^2$ for the Au$_{15}$ clusters as compared to ~$10^4$ GW/cm$^2$ obtained for 140 atom gold-clusters [13] using 100 fs pulses centered at 400 nm. Much smaller threshold intensities for optical limiting in femtosecond regime are known as compared to nanosecond regime [32]. Better optical limiting performance or excited state absorption means that the excited states are more absorbing than the ground state at certain wavelengths and that these excited states have been prepared to have long life times compared to the laser pulse duration [32]. Enhanced nonlinear absorption via photoinduced interstate charge transfer needs much faster rates for electron transfer which has been mostly seen in ns regime [16]. In the femtosecond regime, the laser pulses are very short compared to the time it takes for charge transfer. In Au$_{15}$, the first excited state life time as determined from fluorescence decay time is in the order of 50 ps [10]. The second and higher excited states have much shorter life times. The use of rate equations approximation is not valid in the high intensity ultrafast regime (much shorter than the first excited state life time) [32]. In such a scenario, choosing independently the absorption cross-sections of the ground and the excited states, and the rate parameters in the rate equations for a 'three or five-level model' becomes extremely difficult [16]. Moreover, our case is even more complicated while incorporating the energy transfer via intersystem crossing from Au$_{15}$ to ITO states.

Finally, we can estimate a few useful parameters from the experimentally obtained nonlinear optical coefficients for the gold clusters. The real and imaginary parts of third order optical susceptibility $\chi^{(3)}$ are related to β and γ [33]. These values also have been given in Table 1. The value of β for the Au$_{15}$-ITO system is quite high. For photonic applications such as controlling the amplitude, phase and direction of light of wavelength $\lambda$ in all-optical devices, a figure of merit for the nonlinear materials is defined [34] as F = $\lambda\beta/\gamma$. For the Au$_{15}$-ITO, we estimate F ~ 2 at 395 nm. For practical switching applications, large switching lengths, $L_\pi = \lambda/\gamma I$ are desirous. At the peak intensity of 70 GW/cm$^2$ in our z-scan experiments, we estimate the switching length to be $L_\pi$ ~ 260 μm for the Au$_{15}$-SiO$_2$ and ~90 μm for the Au$_{15}$-ITO samples. Another interesting outcome of our study is the high nonlinear refraction coefficient of the nanoclusters for use as highly nonlinear Kerr medium. For an efficient Kerr material, two parameters $\gamma^2$ and γ/β are sought to be of a high value and the background luminescence due to two photon absorption coefficient β has to be minimal [35]. Previously, single crystal SrTiO$_3$ was reported as one of the most suitable Kerr medium for optical gating application in ultrafast lasers having the $\gamma^2$ ~ 4x10$^{-10}$ cm$^4$/GW$^2$ and γ/β ~ 4.2x10$^{-4}$ cm at 830 nm [35]. For our Au$_{15}$-ITO sample, we estimate $\gamma^2$ ~ 3.6x10$^{-9}$ cm$^4$/GW$^2$ which is better than that for the crystalline SrTiO$_3$, however, the other parameter γ/β is smaller (~2x10$^{-5}$ cm).

## 5. Conclusions

We have studied nonlinear optical response of Au$_{15}$ clusters using z-scan experiments at 395 nm and time-resolved differential transmission measurements. About an order enhancement in the third order optical susceptibility is observed for the Au$_{15}$ clusters in contact with the ITO as compared to the clusters on a SiO$_2$ glass plate. Similar enhancement was observed in the excited state absorption as well. These results indicate a role of the excited state inter-system crossing between the electronic states of the clusters and the ITO to enhance the nonlinear response in the coupled system. We believe that many such coupled systems which are different from the extensively studied polymer-nanoparticle composites will evolve in future to have better understanding of the underlying physics of very small metal clusters in proximity of the tunable planar surface plasmon mode of a metal film.

**Acknowledgements**

We acknowledge financial assistance from Nanomission Project of Department of Science and Technology, Government of India.